\documentclass[11pt,twoside]{article}
\usepackage{asp2010}

\resetcounters

\begin{document}

\title{The Millimeter Astronomy Legacy Team 90 GHz Survey (MALT90) and ALMA}
\author{J.B.~Foster,$^1$ J.M.~Rathborne,$^2$ J.M.~Jackson,$^3$ S.N.~Longmore,$^4$ S.~Whitaker,$^3$ and S.~Hoq$^3$}
\affil{$^1$Department of Astronomy, Yale University, New Haven, CT 06520, USA}
\affil{$^2$CSIRO Astronomy and Space Science, PO Box 76, Epping, NSW 1710, Australia}
\affil{$^3$Institute for Astrophysical Research, Boston University, Boston, MA 02215, USA}
\affil{$^4$European Southern Observatory, Karl-Schwarzschild-Str. 2, 85748 Garching bei Munchen, Germany}

\begin{abstract}
ALMA will revolutionize our understanding of star formation within our galaxy, but before we can use ALMA we need to know where to look. The Millimeter Astronomy Legacy Team 90 GHz (MALT90) Survey is a large international project to map the molecular line emission of over 2,000 dense clumps in the Galactic plane. MALT90 serves as a pathfinder mission for ALMA, providing a large public database of dense molecular clumps associated with high-mass star formation. In this proceedings, we describe the survey parameters and share early science highlights from the survey, including (1) a comparison between galactic and extragalactic star formation relations, (2) chemical trends in MALT90 clumps, (3) the distribution of high-mass star formation in the Milky Way, and (4) a discussion of the ``Brick'', the target of successful ALMA Cycle 0 and Cycle 1 proposals.
\end{abstract}

\section{The MALT90 Survey}
The Millimeter Astronomy Legacy Team Survey at 90 GHz (MALT90) will characterize the physical and chemical conditions of dense molecular clumps associated with high-mass star formation over a wide range of evolutionary states. MALT90 uses the Mopra Spectrometer (MOPS) and the fast mapping capability of the Mopra 22-m radio telescope to map 2000+ candidate dense molecular clumps simultaneously in 16 different lines near 90 GHz. The clumps are drawn from the ATLASGAL \citep{Schuller:2009} survey, and then classified based on their Spitzer morphology to cover a broad range of evolutionary states, from pre-stellar clumps to accreting high-mass protostars and on to H II regions. 

Over the first three years, MALT90 has mapped 1912 dense molecular clumps, which is an order of magnitude more sources than previous comparable surveys \citep[e.g.,][]{Shirley:2003, Pirogov:2003, Gibson:2009, Wu:2010}. This large number of sources allows us to divide the sample into sub-samples (based on mass, evolutionary phase, etc.) while retaining a sufficient number of sources in each sub-sample for statistical analysis. In addition, the large number of sources means that MALT90 includes short-lived phases in the evolution of clumps forming massive stars as well as objects that are intrinsically rare for any other reason. These rare objects will provide interesting targets for followup at higher resolution, primarily with ALMA (the Atacama Large Millimeter Array). 

\begin{figure}[th]
\begin{center}
\includegraphics[width=12cm, angle=0]{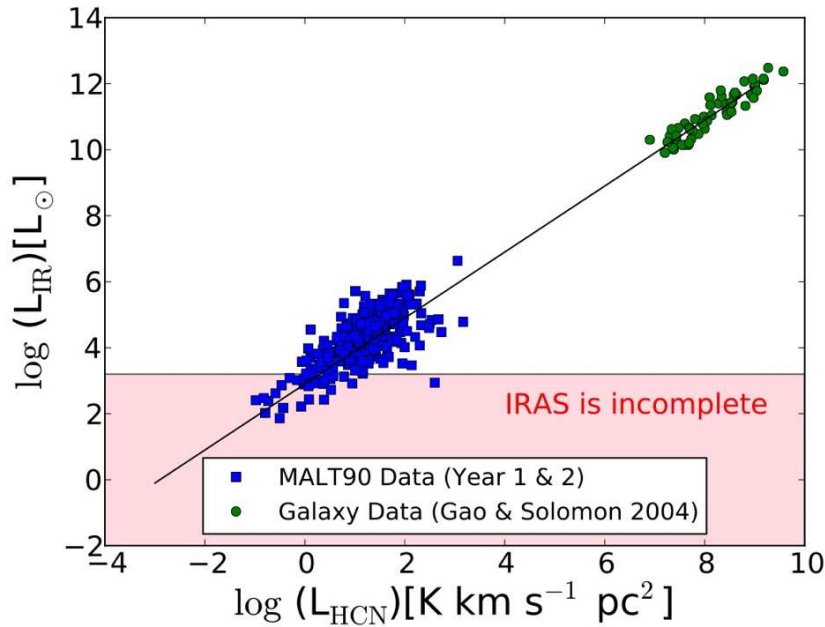}
\caption{Far-infrared luminosity (a proxy for the star formation rate) versus HCN luminosity (a proxy for the surface density of dense gas) for MALT90 sources calculated in the same way as for the external galaxies of \citet{Gao:2004}. The relationship between these two quantities obeys the same relationship in both the MALT90 sources and the external galaxies. }
\label{fig:sfrelations}
\end{center}
\end{figure}

\section{Early Science Highlights}

\subsection{Relationship Between Galactic and Extragalactic Star Formation Relations}
The relationship between gas surface density and star formation rates is a topic of considerable interest for extragalactic studies. Recent work has focused on connecting the relations determined in external galaxies with the same relations within the Milky Way \citep{Kennicutt:2012}. One of the tightest extragalactic star formation relations was described by \citet{Gao:2004}, and connects the luminosity of a galaxy in HCN (a tracer of dense gas) with the far-infrared luminosity (assumed to be a tracer of star formation rate). \citet{Wu:2010} verified that this same relationship holds in massive clumps within the Milky Way. Our much larger sample confirms the results of \citet{Wu:2010}; the \citet{Gao:2004} relationship holds over roughly 10 orders of magnitude to encompass MALT90 clumps (see Figure~\ref{fig:sfrelations}). One important caveat is that the IRAS survey used to estimate the far-infrared luminosity has limited sensitivity, such that a number of MALT90 sources are not detected by IRAS, and are thus not plotted. Investigating the low-luminosity portion of this relationship by using other far-infrared surveys will be future work.

\subsection{Chemical Trends}
\begin{figure}[th]
\begin{center}
\includegraphics[width=12cm, angle=0]{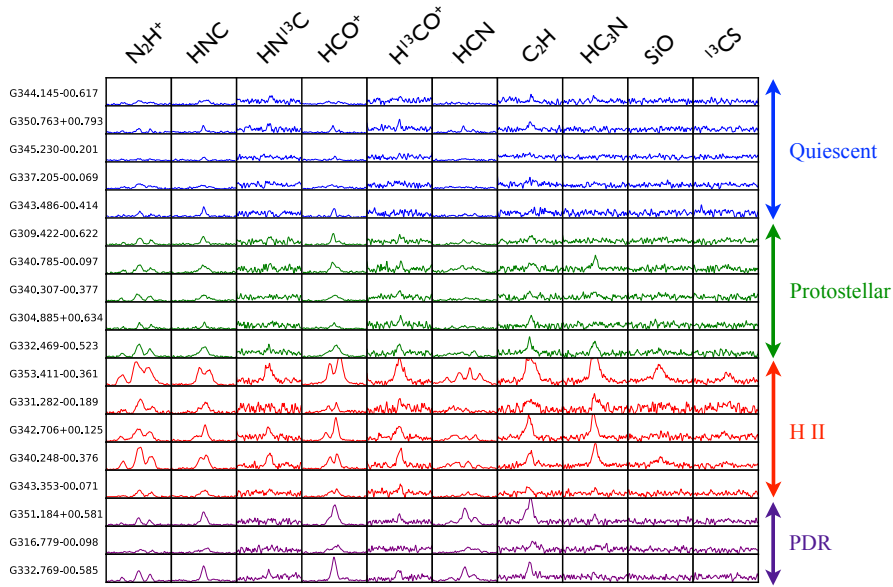}
\caption{Typical spectra for each of the evolutionary states classified in the MALT90 Survey, from the mid-IR dark quiescent clumps (top [blue]) to the protostellar clumps associated with 24 \micron\ point sources (second-from-top [green]), to the bright H II regions (second-to-bottom [red]) and photodisociation regions (PDRs; bottom [purple]). Spectra are shown on the same temperature scale (among classes), and the strongest lines (N$_2$H$^+$, HNC, HCO$^+$, and HCN) are shown scaled down by a factor of 3.5 relative to the other lines. }
\label{fig:spectra}
\end{center}
\end{figure}

We classify MALT90 sources into rough evolutionary states based on their Spitzer morphology; the different evolutionary states show variation in the strengths of molecular lines (see Figure~\ref{fig:spectra}). We have completed a study of the chemical trends in MALT90 clumps observed during the first season \citetext{Hoq et al. submitted}. In particular, we study the ratio of N$_2$H$^+$ to HCO$^{+}$ abundances as a function of evolutionary state of the clumps and the ratio of HCN to HNC integrated intensities. There is no statistically significant trend with evolutionary state in the N$_2$H$^+$ to HCO$^{+}$ abundance ratio, athough models of chemical evolution would predict significant variation as CO (which decreases the abundance of N$_2$H$^+$ relative to HCO$^{+}$) depletes out of the gas phase during pre-stellar collapse and then is released following ignition of the protostar. The HCN to HNC integrated intensity ratio increases as a function of evolutionary state, as expected by models in which this ratio is temperature dependent.

\subsection{Distribution of Massive Star Formation in the Milky Way}
We are able to estimate the distance to all MALT90 sources by using kinematic distances and HI self-absorption \citep[based on SGPS;][]{McClure-Griffiths:2005} to break the kinematic distance ambiguity. Preliminary results from this analysis \citetext{Whitaker et al. in preparation} confirms that MALT90 sources are clustered in previously identified spiral arms of the Milky Way. \citet{Dame:2011} have recently discovered the continuation of the Scutum-Centaurus arm into the first quadrant. In the MALT90 survey region ($l$ $>$ 300) in the fourth quadrant, each line of sight crosses the Scutum-Centaurus arm twice (or is a line of sight down the tangent). This arm is well mapped on the near side of the Galactic center (at a heliocentric distance of 3-4 kpc); a number of MALT90 sources are identified in the far portion of the Scutum-Centaurus arm (heliocentric distances of 15-20 kpc). Future work on MALT90 distances will include using the near-infrared extinction distance methods tested in \citet{Foster:2012} to provide an independent estimate of the distance to each MALT90 source.


\subsection{The Brick}
G0.253$+$0.016, also known as the Brick, is a dark, dense molecular cloud near the Galactic Center. This cloud is potentially the progenitor of a Young Massive Cluster such as the Arches \citep{Longmore:2012}. This object, which is part of the MALT90 survey, was the target for successful ALMA Cycle 0 and Cycle 1 proposals. Preliminary reduction of these data show a wealth of dense filaments with a very complicated velocity structure and complicated chemical patterns  \citetext{Rathborne et al. in preparation}.


\acknowledgements Operation of the Mopra radio telescope is made possible by funding from the National Astronomical Observatory of Japan, the University of New South Wales, the University of Adelaide, and the Commonwealth of Australia through CSIRO.

\bibliographystyle{asp2010}
\bibliography{MyBib2}


\end{document}